\documentclass[aps,pra]{revtex4}

\usepackage{amsfonts}
\usepackage{amsmath}
\usepackage{amssymb}
\usepackage{graphicx}
\usepackage{epsfig}

\newcommand{\wtp}{\widetilde\Psi}

\newcommand{\bp}{\mbox{\boldmath$p$}}

\newcommand{\be}{\mbox{\boldmath$e$}}

\newcommand{\br}{\mbox{\boldmath$r$}}

\begin{document}

\title{Photoionization  accompanied by
excitation at intermediate photon energies}
\author{ E.~G.~Drukarev$^1$,
E. Z. Liverts$^2$,
M.  Ya.  Amusia$^{2,3}$,
R.~Krivec$^4$ and\\
V.~B.~Mandelzweig$^2$\\
\em $^1$ Petersburg Nuclear Physics Institute,\\
\em St. Petersburg, Gatchina
188300, Russia\\
\em $^2$ Racah Institute of Physics,\\
\em The Hebrew University, Jerusalem 91904, Israel\\
\em $^3$ A. F. Ioffe Physical-Technical Institute,\\
\em St. Petersburg 194021, Russia\\
\em $^4$ Department of Theoretical Physics, J.~Stefan~Institute\\
\em P. O. Box 3000, 1001 Ljubljana, Slovenia}

\date{}

\begin{abstract}

We calculate the photoionization with excitation- to
photoionization ratios $R_{n\ell}$ and $R_n=\Sigma_\ell R_{n\ell}$
for atomic helium and positive heliumlike ions at intermediate
values of the photon energies. The final state interactions
between the electrons are included in the lowest order of their
Sommerfeld parameter. This enables us, in contrast to purely
numerical calculations, to investigate the roles of various
mechanisms contributing beyond the high-energy limit. The system
of the two bound electron is described by the functions obtained
by the Correlation Function Hyperspherical Harmonic Method. For
the case of heliumlike ions we present the high energy limits as
power expansion in inverse charge of the nucleus. We analyze the
contribution of excitation of states with nonzero orbital momenta to the
ratios $R_n$. In the case of helium our results for $R_n$ are in
good agreement with those of experiments and of previous
calculations.

\end{abstract}
\maketitle

\section{Introduction}

Atomic photoionization accompanied by excitation and double
photoionization are much studied in experimental and theoretical works
in connection with many-electron problem. In the case of two-electron
systems which is considered in the present paper,
the three-body Coulomb problem is the subject of investigation.

Most attention in experimental  studies  was focused on the
process of double photoionization. Photoionization with additional
excitation have not been investigated in detail. Energy dependence
of the cross section ratios in a broad interval of the photon
energies $\omega$, dependence of these ratios on the value of
nuclear charge $Z$, branching ratios for excitations of ${n\ell}$
subshells of a shell with the principle quantum number $n$ are
still the subjects of future experiments. As it stands now, there
are experimental data only for atomic helium. The intensity of
excitation of $n$-th shell relative to the main photoline $n=1$
was measured in \cite{1} for the photon energies up to several
hundreds eV for the values of $n\le6$. A few measurements of $2s$
and $2p$ excitations at smaller values of the photon energies have
been carried out earlier -- see \cite{1} for references.

Theoretical investigation of the process requires
the knowledge of the wave functions describing two electrons in the field of
the nucleus. In initial state both electrons are bound by the nucleus.
In the final state one of the electrons belongs to continuum, while
the second one is in excited bound state. Certain approximations
(models) for the wave functions are required. Somewhat different
approximations are reasonable in different regions of the photon energy
$\omega$.

We use the terminology which is similar to the one employed for
much studied double photoionization \cite{MG}. It is known that
the ratios
\begin{equation}
R_n(\omega)\ =\ \frac{\sigma^{+*}_n(\omega)}{\sigma^+_1(\omega)}
\end{equation}
of the cross sections $\sigma^{+*}_n(\omega)$ for ionization with
excitation of the second electron to the $n$-th level, to those
without excitation  $\sigma_{1}(\omega)$ do not depend on the
photon energy in {\em high energy limit} \cite{3,DS}
\begin{equation}
R_{n}(\omega)\ =\ \rm const
\end{equation}
for $\omega \rightarrow \infty$.
This requires anyway that $\omega$ exceeds strongly the
values of single-particle ionization potentials $I$
\begin{equation}
\omega\ \gg\ I\ .
\end{equation}

Analysis of \cite{3,DS} have been carried out by employing the
nonrelativistic functions for description of the outgoing
electrons. It was shown in \cite{DK} that asymptotics of the ratio
$R(\omega)$ remains the same in the whole region (3) including the
photon energies corresponding to relativistic outgoing electrons.
Recall that this is not true for the double-to-single
photoionization ratios \cite{AD}.

 By {\em high energies} we mean
that part of the region (3) where the ratios exhibit behavior
described by Eq.~(2). At {\em low energies} the ratio $I/\omega$
cannot be treated as a small parameter.  By {\em intermediate
energies} we mean the values of the photon energies, where
inequality (3) is true, while deviations of the cross section
ratios from the high energy limit are noticeable (with the
relative deviations exceeding 10 percent).  For atomic helium this
is the region from 300--400~eV till 2~keV. In the systems bound by
the nucleus with the charge $Z$ the limits of the interval are
proportional to $Z^2$.

Since the ionization with excitation is a three-body problem, certain
approximated wave functions for both initial and final states are
required. It was shown in \cite{3} and \cite{DS} that in the high
energy limit the final state interactions (FSI) between the electrons
can be neglected. This simplifies the problem of the description of the
final state (under a proper choice of gauge of gauge interactions of
the outgoing electron with the nucleus can be neglected as well). In
\cite{3} the high energy limit of the process was expressed in terms of
the initial state wave function $\Psi_i(\br_1,\br_2)$. The high energy
limits of the ratios $R_n$ for atomic helium were calculated in
\cite{Br,Ab} and dependence on the choice of the approximate function
$\Psi_i$ was traced. The calculations of \cite{AD} include also the $Z$
dependence of the high energy limits of $R_n$. Results of the high
energy calculations for Li$^+$ are presented in \cite{SD} and
\cite{AM}.

At low energies there is no small parameter. All the interactions
involved should be treated as accurately as possible. In this
energy region one must make a choice of both initial and final
state wave functions. The low energy calculations of the cross
sections $\sigma^{+*}_n(\omega)$ have been carried out in
\cite{IB2,PR1} for He and in \cite{IB2, PR2} for Li$^+$. The
paper~\cite{PR2} contains also results for partial cross sections
$\sigma^{+*}_{n\ell}(\omega)$ of ionization accompanied by
excitation of the remaining bound electron to the subshells with
quantum numbers $n$ and $\ell$. Low energy calculations for the
two-electron ions with larger values of $Z$ were carried out in
\cite{IB2}.

In the papers \cite{IB1, IB2} the intermediate energy region was
approached by extension of the low energy calculations to this energy
interval. In the present paper we move from the high energy region by
including next to leading order of expansion in powers of
$\omega^{-1}$. This is achieved by inclusion of the interaction between
the final state electrons in the lowest order of perturbation theory

We find several attractive points in such approach. It provides
the possibility to clarify the role of various mechanisms (in a
fixed form of electron-photon interactions) representing their
contributions in terms of certain characteristics of the initial
wave function. Within the framework of the approach one can
estimate the magnitude of the neglected terms, thus controlling
the accuracy. At the lower limit of the intermediate energy region
numerical calculations with certain models for the final state
wave functions are more precise. The discrepancy between the
results obtained in numerical and perturbative approaches should
diminish as $\omega$ increases. Hence these two approaches should
supplement each other. Similar analysis of the intermediate energy
double photoionization have been carried out earlier \cite{DT}.

We expect the approach developed in the present paper to be useful
also because of certain discrepancies between experimental data
for helium \cite{1} and theoretical results. The high energy limit
of the ratio $R_n$ extrapolated from the data obtained in \cite{1}
is in perfect agreement with the calculated one only for $n=2$.
The disagreement increases with $n$ reaching a factor of about 2
for $n=5$. There is also visible disagreement between theoretical
and experimental results for $R_n(\omega)$ at
$\omega\sim200-400\,$eV for $n=2,4,5$~\cite{IB1,IB2}. Note also
that in the case of helium there is a discrepancy between the
calculations employing various approaches (see, e.g. \cite{IB1}
and \cite{IB2}). Moving from the high energy region can be
instructive also since (as noted in \cite{IB2}) the $R$ matrix
approach, widely used in the low energy calculations becomes
unstable at the high energies. Finally, studies of $Z$ dependence
of the ratios $R_n$ may be of interest in connection with
increasing attention devoted to physics of the multicharged ions.

We calculate ratios (1) of photoionization accompanied by
excitation of the residual ion for  helium atom and light
heliumlike positive ions. We obtain also more detailed
characteristics
\begin{equation}
R_{n\ell}(\omega)\ =\
\frac{\sigma^{+*}_{n\ell}(\omega)}{\sigma^+_{10}(\omega)}
\end{equation}
Such ratios are also detected in the low energy experiments
\cite{W}. The ratios defined by Eq.~(1) can be represented as
$$
R_n(\omega)\ =\sum_\ell R_{n\ell}(\omega)\,.
$$

In this paper the calculations are carried out with inclusion of
next-to-leading order terms of expansion of the ratios (1) in
powers of $\omega^{-1}$. This means that for ${ns}$ states we
calculate the high energy limits of the ratios (1) and the
correction of the order $1/\omega$. For ${n\ell}$ states with
$\ell\ge1$ we obtain the leading order of expansion in $1/\omega$.

In the limit (3) all the interactions of the outgoing electron can
be treated perturbatively \cite{3}. In the high energy limit of
$\sigma_n^{+*}$ final state interactions (FSI) of the outgoing
electron with the electron bound in the residual ion can be
neglected. The excitation following photoionization is due to the
specific correlation in initial state known as {\em shake-up}
(SU). Only $s$ states can be excited by this mechanism.
Excitations of the states with nonzero values of angular momentum
$\ell$ are quenched by a small factor of the order $I/\omega$.

We can present the ratios (4) as
\begin{equation}
R_{ns}(\omega)\  =\ A_n+\frac{I_0}{\omega}B_{n0}
\end{equation}
 (with $I_0$ the electron binding energy in hydrogen) for $\ell=0$, while for
$\ell\ge1$
\begin{equation}
R_{n\ell}(\omega)\ =\ \frac{I_0}{\omega}B_{n\ell}\ ,
\end{equation}
In the atomic system of units used through the paper
($e=m=\hbar=1$, $c=137$) $I_0=1/2$.

The coefficients $A_n$ and $B_{\ell n}$ with $\ell\ge1$ do not
depend on the photon energy, while we show that $B_{n0}$ contains
a smooth dependence on $\omega$. Now we can present the ratios
$R_n$ defined by Eq.~(1) as
\begin{equation}
R_n(\omega)\ =\ A_n+\frac{1}{2\omega}\,B_n; \quad B_n\ =\sum_\ell
B_{n\ell}\,.
\end{equation}

In the SU mechanism the interactions of the outgoing
electron with nucleus can be treated perturbatively \cite{3}.
Excited electrons can be described by the Coulomb field wave functions. Thus,
all the specifics of this three-body problem is contained in the wave
function of the initial state. The ionized electron approaches the
nucleus at the distances which are much smaller than the size of the
atom. The SU cross section is thus determined by initial state  wave
function $\Psi_i(\br_1,\br_2)$ at electron--nucleus coalescence point,
$i.\,e.$ by $\Psi_i(\br_1=0,\br_2)$.

The SU probabilities depend on $n$ in terms of the wave function and of
momentum $p_n$ of the outgoing electron
$$
 p_n^2\ =\ 2\varepsilon_n
$$
with $\varepsilon_n$ being the energy of the outgoing electron.
In the lowest order of expansion
in powers of $I/\omega$ we can put
\begin{equation}
 p_n^2=p_1^2=p^2=2\omega\,.
\end{equation}
The SU terms with this value of $p$ determine the high energy
limit of the ratios
\begin{equation}
\lim_{\omega\to\infty}
R_{n\ell}(\omega)=
\lim_{\omega\to\infty}R_{ns}(\omega)\delta_{\ell0}=A_n\,.
\end{equation}

Now we consider three types of contribution beyond the high energy
limit, in the same way as it was done in \cite{DT} for the double
photoionization. The {\em kinematical corrections} to SU ratios
are caused by taking into account $n$ dependence of momentum $p_n$
in the SU amplitudes. This provides the contributions to the terms
$B_n$ on the RHS of Eq.~(5). Note that these corrections are
proportional to the small parameter $I/\omega$, containing also
dependence on specific parameter $\pi\xi_Z$ with
\begin{equation}
\xi_Z\ =\ \frac{Z}p\ .
\end{equation}
One  has to have in mind that corrections of the order
$\pi\xi_Z/\omega$ drop as $\omega^{-3/2}$ but contain a
numerically large coefficient. We shall not treat $\pi\xi_Z$ as a
small parameter, but include it exactly. The dependence of the
cross sections on $\pi\xi_Z$ known to be presented by the Stobbe
factor $S(\pi \xi)=\exp{(-\pi \xi)}$ which is common for the
photoionization processes \cite{5,6}. These corrections are
expressed in terms of SU contributions $A_n$ to the ratios (5),
which appear only in the ratios $R_{ns}$.

In the next to leading order the excitation energy can be
transferred to the second electron also by the {\em initial state
interactions} (ISI) beyond the SU. In this  case the terms of the
order $1/\omega$ and $Z^2/\omega$ come from the higher terms
$r^2_1/r^2_2$ and $(\br_1\br_2)/r^2_2$ of the expansion of initial
state function $\Psi_i(\br_1,\br_2)$ at $r_1\to0$. Thus the
contribution will be presented in terms of the derivatives of the
initial state wave function, integrated with the Coulomb field
function of the bound state.

The excitation energy can be transferred also by the
{\em final state interactions} (FSI) between the final state electrons.
We include the FSI by perturbative method developed
in \cite{7}. The FSI amplitude is presented as power series of the Sommerfeld
parameter of the interaction between the fast outgoing electron and
that of the residual ion
\begin{equation}
\xi\ =\ \frac 1 v\ ,
\end{equation}
while $v$ is their relative velocity. Thus, the square of the
amplitude is presented as power series in $\xi^2=1/2\varepsilon$
with $\varepsilon$ being the energy of the outgoing electron.
Looking for the terms of the relative order $\omega^{-1}$ in the
cross sections, we must include the lowest correction of the order
$\xi^2$, putting
\begin{equation}
\xi^2\ =\ \frac{1}{2\omega}\,.
\end{equation}
The FSI contributions are presented in terms of matrix elements of
relatively simple operators sandwiched by the function $\Psi_i(
\br_1, \br_2)$ and the Coulomb function of the electron in
residual ion. The states with any angular momenta $\ell$ can be
excited by the FSI in the next to leading order of $\omega^{-1}$
expansion.

Thus, all the contributions up to the order $\omega^{-1}$ will be
presented in terms of certain characteristics of the initial state
wave functions. We employ the functions obtained by Correlation
Function Hyperspherical Harmonic Method (CFHHM), obtained in
\cite{8}. The CFHHM functions have been employed successfully for
investigation of the parameters of the bound two-electron
systems \cite{9} and of some characteristics of the double
photoionization. Also, the method of inclusion the FSI \cite{7}
enabled earlier to remove the discrepancy between experimental and
theoretical results in creation of vacancies in electronic shells
during nuclear transitions and in single photoionization of the $p$
states. In \cite{DT} it was used for investigation of the double
photoionization at intermediate energies. In the present paper we use
the CFHHM functions and the perturbative treatment of FSI for
investigation of photoionization with excitation.

Note that for the system containing larger number of electrons the
picture is more complicated. Considering ionization with
excitation of the subshell with $\ell=1$ we find for ionization
without excitations $\sigma^+ \sim \omega^{-9/2}$, while for
ionization accompanied by excitation to an $s$ state $\sigma^{+*}
\sim \omega^{-7/2}$ (the ISI provides admixture of two $s$ state
electrons to the system containing two $p$ state ones). Hence the
corresponding ratio increases proportionally to  $\omega$.

Our analysis is completely nonrelativistic. We neglect the terms
of the order $\omega/m$ in the wave function of the  final state,
and in the operator of the photon--electron interaction. The
latter means that we are using the dipole approximation. We assume
also $(Z/137)^2\ll 1$, to neglect relativistic effects in the
initial bound system.

\section{General equations}

The cross section of photoionization accompanied by excitation of the
residual ion into a state with the quantum numbers $n,\ell,m$ can be written as
\begin{equation}
d\sigma^{+*}_{n\ell}\ =\ \frac {2\pi}{\omega c}
\sum_m|\overline{F_{n\ell m}}|^2
\delta(\omega-\varepsilon_n-I_i)\,\frac{d^3p_n}{(2\pi)^3}\,.
\end{equation}
Here $I_i$ denotes the ionization potential of K electron in initial state
atom. The factor 2 is due to two electrons in the K shell. The overline
shows that the averaging over the photon polarizations have been
carried out.
The angular dependence of the amplitudes can be written explicitly due
to the dipole approximation employed
The amplitude
\begin{equation}
F_{n\ell m}\ =\ \langle\Phi_{n\ell m}|\gamma|\Psi\rangle\,,
\end{equation}
with $\gamma$ being the operator of interaction between the photon and an
electron, while $\Psi$ and $\Phi_{n\ell m}$ describe the initial and final
two-electron states, can be represented as
\begin{equation}
F_{n\ell m}\ =\ (4\pi)^{1/2}\frac{(\be \cdot \bp_n)}c\ T_{n\ell
m}\,.
\end{equation}
After averaging over the photon polarizations one obtains
\begin{equation}
\sigma^{+*}_{n\ell}\ =\
\frac4{3}\cdot\frac{p^3_n}{c^3\omega}\sum_m |T_{n\ell m}|^2\,.
\end{equation}

If the FSI are neglected, the final state function
is
\begin{equation}
\Phi_{n\ell m}(\br_1,\br_2)\ =\ \psi_f(\bp_n;\br_1)
\psi_{n\ell m}(\br_2)
\end{equation}
with the functions $\psi_f$ and $\psi_{n,\ell,m}$ being just the
continuum and bound state single-particle wave functions in the
Coulomb field. If condition (3) is valid, the interactions of the
outgoing electron with the nucleus can be included perturbatively.
Using the velocity gauge for the operator $\gamma$, i.e.
$$
\gamma(r)\ =\ -i(\be\cdot\mbox{\boldmath$\nabla$})
$$
with $\be$ standing for the photon polarization,
 we
obtain the leading contribution of expansion in powers of $p^{-2}$
as coming from the plane waves. Following \cite{DT} we can
separate two scales in the interactions between the fast outgoing
electron and the nucleus. Those taking place at small distances of
the order $p^{-1} \ll r_c$ with $r_c=1/Z$ being the characteristic
size of the atom are expressed in terms of the parameter
$\pi\xi^{(n)}_Z$ $(\xi^{(n)}_Z = Z/p^{(n)}$. Such contributions
can be calculated explicitly, producing the factor
\begin{equation}
N(\xi^{(n)}_Z)\ =\ N_r(\xi^{(n)}_Z)\,e^{-\pi\xi^{(n)}_Z}
\end{equation}
with $N_r(\xi^{(n)}_Z)=\left(2\pi\xi^{(n)}_Z/(1-\exp{(-2\pi
\xi^{(n)}_Z}))\right)^{1/2}$ being the normalization factor of the
nonrelativistic Coulomb continuum wave function. The interactions
which take place at the distances of the order $r\sim r_c$ can be
presented as $p^{-2}$ series thus cancelling in the ratios (1) and
(2). Thus we can put
\begin{equation}
\Phi^{(0)}_{n\ell m}(\br_1,\br_2)\ =\ N(\xi^{(n)}_Z)\,
e^{i(\bp_n\br)}\psi_{n\ell m}(\br_2)\,.
\end{equation}
Following \cite{8} we present the factors $N^2(\xi_Z^{(n)})$ in the
cross sections as
\begin{equation}
 N^2\left(\xi_Z^{(n)}\right) =\ h(\pi\xi_Z^{(n)})e^{-\pi\xi_Z^{(n)}}
\end{equation}
with the function $h(\xi_Z^{(n)})=2\pi \xi_Z^{(n)}/(\exp{(\pi
\xi_Z^{(n)})}+\exp{(-\pi \xi_Z^{(n)})})$ containing only weak
dependence on parameter $\pi\xi^{(n)}_Z$. Thus we can put
$h(\pi\xi_Z^{(n)})=h(\pi\xi_Z)$, with $\xi_Z$ defined by Eq.~(10).
Hence,
\begin{equation}
 N^2\left(\xi_Z^{(n)}\right) =\ h(\pi\xi_Z)e^{-\pi\xi_Z^{(n)}}\,.
\end{equation}
The second factor on the RHS of Eq.~(21) is the Stobbe factor mentioned
 above.

We shall present the perturbative FSI contributions also in terms of
the function (17). Thus the ratios (1) will be presented in terms of the
matrix elements of initial state two-electron function and the Coulomb
function of the excited electron.

\section{Amplitudes beyond the shake-up}

Following the analysis given above, we present the amplitudes for
ionization with excitation beyond the SU as
\begin{equation}
F_{n\ell m}\ =\ F_{n\ell m}^{(s)}+F_{n\ell m}^{(i)}+F_{n\ell m}^{(f)}
\end{equation}
with $F_{n\ell m}^{(s)}$ standing for SU amplitude, which includes
kinematical corrections to the high energy limit, while $F_{n\ell
m}^{(i)}$ and $F_{n\ell m}^{(f)}$ are the contributions caused by
correlations in ISI and FSI correspondingly.

\subsection{Amplitudes without inclusion of final state interactions}

Since in this subsection we neglect the interactions between the
outgoing electrons, we can use Eq.~(14) for the amplitude with the
final state wave function presented by Eq.~(17). Recall that we
use the operator $\gamma$ in the velocity form. This provides
\begin{equation}
F_{n\ell m}\ =\ (4\pi)^{1/2}(\be\cdot\bp_n)\,N(\xi^{(n)}_Z) \int
d^3r_2\psi^*_{n\ell m}(\br_2)\wtp_i(\bp_n;\br_2).
\end{equation}
Here
\begin{equation}
\wtp_i(\bp_n;\br_2)\ =\int d^3r_1\Psi_i(\br_1,\br_2)\,
 e^{-i(\bp_n \cdot\br_1)}
\end{equation}
is the partial Fourier transform of the initial state wave function in
variable $\br_1$.

Since the integral over $r_2$ on the RHS of Eq.~(23) is saturated
at $r_2\sim r_c$, while $p_n\gg r^{-1}_c$, we need expansion of
the function $\wtp(\bp_n;\br_2)$ in inverse powers of $p^2_n$. It
is convenient to employ the Lippman--Schwinger equation
\begin{equation}
\wtp_i(\bp_n;\br_2)= \frac{2Z}{p^2_n}\,J(\bp_n,\br_2); \quad
J(\bp_n,\br_2)=\int\frac{d^3r}r\,e^{-i(\bp_n\cdot
\br)}\Psi_i(\br,\br_2) .
\end{equation}
The integral on the RHS is dominated by $r\sim p^{-1}_n\ll r_2$. Thus
the expansion in $p^{-2}_n$ can be carried out by expanding the
function $\Psi(\br,\br_2)$ in powers of $r$ in the integrand
on the RHS of Eq.~(25).

\subsubsection{Shake-up with kinematical corrections}

Presenting
\begin{equation}
J(\bp_n,\br_2)\ =\ \lim\limits_{\delta\to0} \int \frac{d^3r}r\,
e^{-i(\bp_n\br)-\delta r}\Psi_i(\br,\br_2),
\end{equation}
we obtain for the leading order contribution
\begin{equation}
J(\bp_n,\br_2)\ =\ \frac{4\pi}{p^2_n}\ \Psi(0,r_2)\ ,
\end{equation}
which enable to write for the SU amplitudes
\begin{equation}
F^{(s)}_{n\ell m}= a(p_n)S_n\delta_{\ell0}\delta_{m0}; \quad S_n=(4\pi)^{1/2}
\int dr_2r^2_2\psi^{(r)}_{n0}(r_2)\Psi(0,r_2)
\end{equation}
with
\begin{equation}
a(p_n)\ =\ (\be\bp_n)N (\pi\xi^{(n)}_Z)\, \frac{8\pi
Z}{p^2p^2_n}\,,
\end{equation}
while the upper index $(r)$ in Eq.~(28) labels the radial part of the
 Coulomb function $\psi_{n0}$. In the leading order we should neglect
the dependence of $p_n$ on $n$, putting $p_n=p$, just as in
Eq.~(8). The high energy limit of the amplitude (28) is thus
\begin{equation}
F^{(0)}_{n00}\ =\ a(p)S_n\,.
\end{equation}
In the next to leading term we must include the $n$ dependence of
$p_n$. Since the residual ion contains only one electron, the
latter is described by the Coulomb wave function, and thus
\begin{equation}
p_n^2\ =\ p^2-2\delta_n\,,
\end{equation}
where
\begin{equation}
\delta_n=\frac{Z^2}{2}(1-\frac{1}{n^2})
\end{equation}
is the excitation energy of the electron in the final state ion.

\subsubsection{Contributions of correlations in the initial state}

Now we return to Eqs.(24), (25), looking for higher order terms of
expansion of the function $\Psi_i(\br,\br_2)$ at $r \rightarrow
0$. Since the CFHHN functions are expressed in terms of variables
$r=|\br|,
 r_2=|\br_2|, \rho=|\br-\br_2|$, we present the expansion in terms of
the function $\Psi(r,r_2,\rho)=\Psi_i(\br,\br_2)$
\begin{equation}
\Psi(r,r_2,\rho)=(1+ r_i\nabla_i+\frac{1}{2}r_ir_j\nabla_i\nabla_j)
\Psi(r,r_2,|\br-\br_2|).
\end{equation}
Here we put $\Psi(r, r_2, |\br-\br_2|)=\Psi(0,r_2,r_2)$ after the derivatives
are calculated. Using Eq.~(33) we find that at small $r$
\begin{eqnarray}
&& \hspace*{-0.5cm}
\Psi(r,r_2,\rho)=\Psi(0,r_2,r_2)+ r\Psi^{'}_r(r,r_2,r_2)-
r\tau \Psi^{'}_{\rho}(0,r_2,\rho)+
\nonumber \\
&&+\ \frac{r^2}{2}\Psi^{''}_r(r,r_2,r_2)
+\frac{r^2(1-\tau^2)}{r_2}\Psi^{'}_{\rho}(0,r_2,\rho)+
\nonumber\\
&&+\ \frac{r^2\tau^2}{2}\Psi^{''}_{\rho}(0,r_2,\rho)-
r^2\tau\Psi^{'}_{r\rho}(r,r_2,\rho).
\end{eqnarray}
Here $\tau=(\br \cdot \br_2)/r\cdot r_2$. The derivatives
$\Psi'_r$ and $\Psi'_{\rho}$ (and those of the second order) are
taken at the points $r=0$ and $\rho=r_2$. The higher terms of
expansion in $r$ contribute to the higher order corrections in
$1/p$ to the amplitudes. Thus, they are neglected. While
evaluating the next to leading order terms we must put $p_n=p$.

Using Eqs. (26) and (34) we find nonzero contributions to the
amplitudes with the angular momenta $\ell=0$ and $\ell=1$. For
excitation to $s$ states we obtain
\begin{eqnarray}
F^{(2)}_{n00}&=&a(p)Q_n\frac{1}{p^2};
\nonumber \\
 Q_n&=&-(4\pi)^{1/2}
\int dr_2r^2_2\psi^{(r)}_{n0}(r_2)\ \times
\nonumber\\
&& \times\ \bigg[
\Psi_r''(r,r_2,r_2)+\frac{1}{3}\Psi_{\rho}''(0,r_2,\rho)+
\frac{2}{3r_2}\Psi_{\rho}'(0,r_2,\rho)\bigg]r^2_0
\end{eqnarray}
with the upper index $(r)$ labelling the radial part of the
single-particle Coulomb field function as in Eq.~(28). The
function $a(p)$ is determined by Eq.~(29). Other notations are
explained in the text below Eq.~(26). For excitation into $p$
states, choosing the direction of the outgoing electron momentum
as the axis of quantization of the angular momentum, we obtain
\begin{equation}
F^{(2)}_{n1m}=ia(p)P_n\frac{\delta_{m0}}{p}; \quad
P_n=(4\pi)^{1/2}\,\frac{2\sqrt3}3 \int
dr_2r^2_2\psi^{(r)}_{n1}(r_2) \Psi'_\rho(0,r_2,\rho)r_0\,.
\end{equation}

Thus interactions in the initial state provide corrections of the
order $p^{-2}$ to the cross sections of excitations into $s$
states. The dependence of the wave function on the interelectron
distances $\rho=|\br_1-\br_2|$, which describes the electron
correlations enables also excitations into $p$ states. Excitation
to the states with higher orbital momenta due to the ground state
correlations only are still impossible.

\subsection{Contribution of the final state interactions}

Now we include the final state interactions. Following \cite{7,10} we
present the final state wave function as
\begin{equation}
\Phi^{(f)}\ =\ (1+GV_{ee}+GV_{ee}GV_{ee})\Phi^{(0)}
\end{equation}
with $\Phi^{(0)}$ being the wave function (17), where the FSI have
been neglected (here we omit lower indices), $V_{ee}$ is the
electron--electron interactions, $G$ is the propagator of the
system of two non-interacting electrons in the Coulomb field of
the nucleus. The second and third terms on the RHS of Eq.~(37)
correspond to one and two interactions between the final state
electrons, thus being proportional to the powers of the parameter
$\xi$ defined by Eq.~(11).

The two last terms on the RHS contain infrared divergent
contributions caused by the Coulomb interactions $V_{ee}$. It was
shown in \cite{7} that the infrared divergent terms cancel in each
order of the expansion of the square of the amplitude $|F|^2$. The
situation is similar to that with the infrared singularities in
the $e-N$ scattering analyzed in \cite{11}. The cancellation can
be illustrated by assuming the electron interactions to be defined
as $V_{ee}(r)=\lim_{\nu\to0}e^{-\nu r}/r$. The contributions
$\ln\nu$ emerge in intermediate steps but vanish in the final
expression for $|F|^2$.

Explicit expressions which include the FSI in process with the fast
outgoing electron up to the terms of the order $\xi^2$ have been
obtained in \cite{7,10}. The first order amplitude $F^{(f1)}$,
corresponding to the second term on the RHS of Eq.~(37) is mostly
imaginary. The real part of $F^{(f1)}$ is suppressed by additional power
of $p^{-1}$ and thus can be written as being proportional to
$\xi^{-1}$. The second order amplitude is mostly real. Thus
Im$\,F^{(f1)}\sim\xi$, Re$\,F^{(f1)}\sim\xi^2$, Re$\,F^{(f2)}\sim\xi^2$,
Im$\,F^{(f2)}\sim\xi^3$ (we do not trace the dependence on $Z$ here).
The FSI amplitudes can be presented as \cite{7}
\begin{eqnarray}
F^{(f1)}_{n\ell m} &=&a(p)\left(i\xi\langle\psi_{n\ell m}|\ln(r_2-r_{2z})
\nu|\Psi_i\rangle+\frac{\xi^2}2\langle\psi_{n\ell m}|r_0\frac d{dr_2}
|\Psi_i\rangle\right);\\
F^{(f2)}_{n\ell m} &=&-\frac{a(p)\xi^2}2\ \langle\psi_{n\ell m}|\ln^2
(r_2-r_{2z})\nu|\Psi_i\rangle\,.
\end{eqnarray}
Here $\Psi_i \equiv \Psi_i(r_1=0,r_2)$, is a function of
$r_2=|\br_2|$, $z$ is the direction of the momentum of the
outgoing electron. Recall that $z$ axis is chosen for the
quantization of angular momentum, $r_0$ is the Bohr radius. Thus
all the contributions to $F^{(f)}_{n\ell m}$ have nonzero values
only for $m=0$.

For $s$ states both amplitudes $F^{f1}$ and $F^{f2}$ are important,
since the terms containing the factor $\xi^2$
interfere with the SU amplitude. For $s$ states
\begin{equation}
F^{(f1)}_{ns}=a(p)(i\xi U_n+\xi^2V_n) ; \quad
F^{(f2)}_{ns}=-a(p)\xi^2W_n,
\end{equation}
with
\begin{eqnarray}
U_n &=& (4\pi)^{1/2}\int dr_2r^2_2\psi^{(r)}_{n0}(r_2)\ln
r_2\nu\,\Psi_i(0,r_2);
\nonumber \\
V_n &=& \frac{(4\pi)^{1/2}}{2}\int dr_2r^2_2\psi^{(r)}_{n0}(r_2)
\frac{d\Psi_i(0,r_2)}{dr_2}\,r_0;
\nonumber\\
W_n &=& -\frac{(4\pi)^{1/2}}{2}\int dr_2r^2_2\psi^{(r)}_{n0}(r_2)\ln^2(r_2\nu)
\Psi_i(0,r_2).
\end{eqnarray}

For the states with $\ell\neq0$ there is no interference
with the SU amplitude. Thus only the first term of the
amplitude $F^{(f1)}$ is important.
We can present
\begin{equation}
F^{(f1)}_{n\ell}\ =\ ia\xi S_{n\ell}c_\ell +0(\xi^2)
\end{equation}
with
\begin{equation}
S_{n\ell}\ =\ (4\pi)^{1/2}\int dr_2r^2_2
\psi^{(r)}_{n\ell}(r_2)\Psi(0,r_2)\,,
\end{equation}
while
\begin{equation}
c_\ell\ =\ \frac{\sqrt{2\ell+1}}2\int\limits^1_{-1} dt\ln(1-t){\cal
P}_\ell (t)=-\frac{\sqrt{2\ell+1}}{\ell(\ell+1)}\,,
\end{equation}
and ${\cal P}_\ell$ is the Legendre polynomial.

\section{The ratios}

Now we can calculate the cross sections and the ratios (1) and
(4). The cross sections are related to the squares of the
amplitudes $|F|^2$ by Eqs. (13) and (16). We start with
calculation of  the values of $|F|^2$.

\subsection{Excitation of $s$ states}

Expressions for excitation of $s$ states have the most complicated
structure
\begin{equation}
|F_{ns}|^2 = a^2(p_n)S_n^2
+\frac{a^2(p)}{2\omega}\left[2S_nQ_n+2S_n(V_n+W_n)+U_n^2\right].
\end{equation}
Here the  first term on the RHS stands for SU contribution with
account of kinematical corrections -- Eqs. (28) and (29). The
first term in the square

brackets comes from interference between SU and ISI amplitudes --
Eqs. (30) and (35). The second term in brackets is caused by
interference of SU amplitude presented by Eq.~(30) with the first
and second order FSI amplitudes presented by Eqs. (40) and (41).
The last term in brackets is a purely FSI contribution.

In order to obtain contribution of the first term on the RHS of
Eq.~(45) to the ratio (1) we must include the $n$ dependence of
the phase volume in Eq.~(13) for the cross section. As a result,
for purely SU ratio we find
\begin{equation}
R_{ns}^{SU}\ =\ \frac{S_n^2}{S_1^2}
\frac{p}{p_n}e^{-\pi(\xi^{(n)}_Z-\xi)},
\end{equation}
with $p_n$ defined by Eq.~(31). In the lowest order of expansion
in powers of $I/\omega$ the dependence of $R_n^{SU}$ on parameter
$\pi\xi_Z$ is just the same as it would result in expansion of the
RHS of Eq.~(46) in powers of $\pi\xi_Z$. However this is not true
for the higher order terms of $I/\omega$ expansion.

The contribution of the other terms to the ratios $R_n$ can be
found as their ratios to squared amplitude of photoionization
without excitation $|F_{1s}|^2$, where the corrections of the
order $I/\omega$ also should be included. This gives
\begin{eqnarray}
&& \hspace*{-0.5cm}
R_{ns}(\omega)=\frac{S_n^2}{S_1^2}+\frac{1}{2\omega}\bigg[
\frac{S_n^2}{S_1^2} \frac{Z^2}2(1-\frac1{n^2})(1-\pi\xi_Z)+
\nonumber \\
&&+\ \frac{2S_n}{S_1^2}\left(Q_n +V_n+W_n-\frac{S_n}{S_1}
(Q_1+V_1+W_1)\right)+U_n^2-\frac{S_n^2}{S_1^2}U_1^2 \bigg].
\end{eqnarray}
The first term on the RHS is the high energy limit of the ratio.
Note that Eq.~(47) provides exact dependence on parameter
$\pi\xi_Z$ in next to leading order of $I/\omega$ expansion.
Expression in square brackets on the RHS of Eq.~(47) should be
identified with parameter $B_{n0}$ introduced by Eq.~(5).
Separating energy independent contributions and the terms, which
depend on the photon energy through parameter $\pi\xi_Z$ we write
\begin{equation}
B_{n0}\ =\ d_n+\pi\xi_Zf_n\,,
\end{equation}
 with
\begin{eqnarray}
d_n &=& \frac{S^2_n}{S^2_1} \frac{Z^2}2(1-\frac1{n^2})
+\frac{2S_n}{S_1^2}\left(Q_n +V_n+W_n-\frac{S_n}{S_1}
(Q_1+V_1+W_1)\right)+U_n^2-\frac{S_n^2}{S_1^2}U_1^2\,;
\nonumber \\
f_n &=& -\frac{S_n^2}{S_1^2}\frac{Z^2}2\left(1-\frac1{n^2}\right).
\end{eqnarray}

\subsection{Excitation of \boldmath$p$ states}

Excitations of the states with $\ell\neq0$ can take place only
beyond the SU approximation. The contributions of  ISI and FSI are
expressed by Eq.~(36) and by the first term in brackets on the RHS
of Eq.~(38) correspondingly.
\begin{equation}
F_{n1}\ =\ i\,\frac{1}p(P_n+S_{n1}),
\end{equation}
leading to
\begin{equation}
R_{n1}(\omega)\ =\ \frac{1}{2\omega}\
\frac{(P_n+S_{n1})^2}{S_1^2}\,.
\end{equation}

\subsection{Excitation of the states with \boldmath$\ell\ge2$}

In this case  only the FSI contribute. The amplitude is expressed
by  Eq.~(43) giving
\begin{equation}
R_{n\ell}(\omega)\ =\ \frac{1}{2\omega}\ \frac{c^2_\ell
S^2_{n\ell}}{S_1^2}
\end{equation}
with $c_\ell$ defined by Eq.~(44).

Now we turn to analysis of particular cases.

\section{The case of Helium}

\subsection{High energy limit}

In Table I we compare our results for the high energy limits with
those extrapolated from the experimental data \cite{1} and with
the results of previous calculations \cite{Br,SD,IB1,IB2}.  There
were also several publications of the high energy limit for
$R_{2s}$ only. The pioneering calculation \cite{16} provided
$R_{2s}=4.61\cdot 10^{-2}$ , while the latest available result is
$R_{2s}=4.79 \cdot 10^{-2}$ \cite{17}.

One can see that various theoretical approaches provide very close
values for high energy limits at $n\le6$. However the values of
these limits extrapolated in \cite{1} from the experimental data
are in perfect agreement with the theoretical results only for
$n=2$. Discrepancy between theoretical and experimental results
increases with $n$ rapidly.

\subsection{Beyond the high energy limit}

Now we consider the contributions beyond the limit (9). Let us
start with excitation of $ns$ states. As we showed above, the
contributions beyond the high energy limits come from kinematical
corrections to the SU term and from the initial and final states
electron--electron interactions. One can see from Table~2 that the
FSI and ISI contributions are positive, with the FSI term being
about 3 times larger than the ISI one for all values of $n$. The
kinematical corrections are negative at $\mu>1$, corresponding to
$\omega<540\,$eV. At larger $\omega$ values they become positive.
Their contribution to the parameter $B_{n0}$ defined by Eq.~(5)
does not exceed 10 percent for $\omega\lesssim1\,$keV. They
increase  with $\omega$, becoming as large as  25-30 \% in the
limit $\mu\ll1$ ($\omega\gg500\,$eV). In this limit the FSI
contribution determines about one half of the parameter $B_{n0}$.
Note that the relative role of the three contributions does not
vary much with $n$.

Contributions to excitation of $p$ states come from FSI and ISI.
Actually the former one dominates, providing more than 4/5 of the
total contribution to $B_{n1}$. The result of calculations are
presented in Table~3. One can see that $B_{n0}$ and $B_{n1}$
provide contributions of the same order of magnitude to the energy
dependent part of the ratio $R_n(\omega)$ defined by Eqs. (1) and
(7).

Thus the energy dependent parts of the ratios $R_n(\omega)$
determined by parameters $B_n$, are dominated by contributions of
$s$ and $p$ states. The $d$ states provide corrections of about
10\%, while contributions of states with larger values of orbital
momenta are negligibly small. The coefficients $B_n$ are dominated
by FSI which provide more than 70\% of the values.

Excitation of the states with $\ell\ge2$ are due to FSI only. The
cross sections of $d$ states excitations are several  times
smaller than those of $p$ states still providing noticeable
contributions to parameter $B_n$ (Eq.~(71)) for $n\ge3$. The
relative role of excitation of $d$ states slowly increases with
increasing $n$ -- see Table~3. Excitation of states with higher
values of $\ell$ drops rapidly with increasing of $\ell$.  For
example, the cross section for excitation of $4f$ state is about
twenty times smaller than that of $4d$ state. The values of
$B_{n\ell}$ and $B_n$ are presented in Table~3.

Energy dependence of the relative role of excitations to the
states with $\ell>0$,expressed by the ratio $\sigma^{+*}_{n
\ell}/\sigma^{+*}=R_{n\ell}/R_n$ is shown in Fig.~1.

The ratios $B_n/A_n$ converge
to certain limiting values while $n$ increases. This is due to similar
$n^{-3}$ behavior of both characteristics at large $n$.

\subsection{Comparison with earlier results}

Now we compare our result with experimental and theoretical data
obtained by the others. In Fig.~2 we present the ratios $R_n$ for $2\le
n\le6$, calculated in the present work and measured in Ref.~\cite{1}.
We show also the results of calculations carried out in \cite{IB2},
where the intermediate energy region was reached by moving from the low
energies.

We see that for $n=2$ our results are in good agreement with those
of \cite{1,IB2} for $\omega\ge400\,$eV. As expected, there are
noticeable deviations from results of \cite{1,IB2} at smaller
values of $\omega$. We do not show the results of calculations for
$\omega\sim900\,$eV obtained in \cite{TB,17}, which are also in
good agreement with those of the present paper. For $n=3$ we find
a good agreement with experimental and theoretical results at all
values of $\omega$. For the cases $n=4$ and $n=5$ our results are
close to those of \cite{IB2}, with both sets of the calculated
values exceeding the experimental data at $\omega\sim300-400\,$eV.
For $n=6$ the experimental results are available only for
$\omega\le160\,$eV, where the accuracy of our approach is poor
since $\xi^2_Z\ge0.34$. However the deviations between our results
and experimental data is are not large for $n=6$, as well as for
the other values of $n$ in this energy region.

\section{\boldmath$Z$ dependence}

\subsection{High energy limit}

It was shown in \cite{AD} that the SU ratios $A_{n}$ drop as $Z^{-2}$
at $Z\gg1$. The tendency is illustrated by the results of calculations
for $Z\le10$ presented in Table~4. One can see that the convergence
to $Z^{-2}$ law becomes better for larger values of $n$. In \cite{19}
 $Z$ dependence of high energy limits for double-\-to-\-single photoionization
ratio was traced and presented as a $Z^{-1}$ series. We can write
similar presentation for the ionization followed by excitation.
Assuming that $A_n$ can be approximated by two terms of the series
one has
\begin{equation}
A_n\ =\ \frac{a_n}{Z^2}+\frac{b_n}{Z^3}\,.
\end{equation}
The values of $a_n$ and $b_n$ are given in Table 5. The
convergence of $Z^{-1}$ series is faster than in the case of
double photoionization \cite{19}. Also, in contrast to the double
photoionization case, the  leading $Z^{-2}$ terms underestimate
the values of the ratios. Note that in approach of \cite{AD}, where all
the interactions between the electrons where treated perturbatively,
\begin{equation}
A_n\ =\ \frac{c_n}{Z^2}
\end{equation}
with $c_n$ also presented in Table~5 (a numerical error was corrected
in \cite{AM}). As expected, our values of $a_n$ are close to $c_n$.

The results presented in Table 5 illustrate  also the tendency to
$n^{-3}$ behavior. The values of the product $n^3\cdot
R^{(0)}_{ns}$ for $n=5$ and $n=6$ differ by $5\%$ for $Z=2$ and by
$4\%$ for $Z=10$.

\subsection{Beyond the high energy limits}

In this case Eq. (3) can be written as
$$
\omega\ \gg\ \frac{Z^2}2\,.
$$
In order to trace $Z$ dependence of the characteristics, we
consider the limit $Z\gg1$. Let us start with excitations of
$s$-states. One can see that in Eqs. (48) and (49) the ratio
$s_1/S_n\sim1/Z$, while $Q_n/S_1\sim Z$, $V_n/Z\sim Z$,
$W_n/S_1\sim1$, $U_n/S_1\sim1$. Thus in Eq.~(5) $A_n\sim1/Z^2$,
$B_{ns}\sim1$. Hence, we can present
\begin{equation}
R_{n0}\ =\left(1+\frac{Z^2}{2\omega}r_{n0}\right)A_n\,,
\end{equation}
with $r_{n0}=B_{n0}/Z^2A_n$. Using Eqs.~(48) and (49) we can write
\begin{equation}
r_{n0}\ =\ r^d_{n0}+\pi\xi_Z r^f_{n0}\,
\end{equation}
with $r^d_{n0}=d_n/Z^2A_n$ and
\begin{equation}
r^f_{n0}=\frac{f_n}{Z^2A_n}=-\frac{n^2-1}{2n^2}\,.
\end{equation}

In similar way we can present for $\ell\ge1$
\begin{equation}
R_{n\ell}\ =\ \frac{Z^2}{2\omega}r_{n\ell}A_n\,,
\end{equation}
with $r_{n\ell}=B_{n\ell}/Z^2A_n$, while parameters $B_{n\ell}$
are introduced by Eq.~(6).


For the cross section ratios $R_n$ defined by Eq.~(1) we present
\begin{equation}
R_n=\left(1+\frac{Z^2}{2\omega}\,r_n\right)A_n\,; \quad
r_n=r^d_n+\pi \xi_Z r^f_n ; \quad r^d_n=r^d_0+\sum_{\ell>0}
r_{n\ell}\, ; \quad r^f_n=r^f_{n0}\,.
\end{equation}

For illustration we present characteristics of the process for
$Z=10$. Interplay of the three types of contributions to the
parameter $B_{n0}$ describing excitation of $s$ states is shown in
Table 6. As well as in the case $Z=2$ the FSI provides the main
contribution. However the domination is less pronounced than in
the case of helium.

As one can see from Eq.~(57), the ratios $r^f_{n0}$ do not depend
on $Z$. Dependence of parameters $r_{n\ell}$ and $r_n$ on $Z$ is
illustrated by the results presented in Table~7.  The values of
$r^d_n$ for $Z=10$ are somewhat larger than for $Z=2$. This is
mainly due to larger contribution for excitation of $p$ states in
the case $Z=10$. On the other hand the role of excitation of $d$
states becomes smaller -- see Fig.~3. The ratio of $r^d_n$ for $Z=2$
and $Z=10$ exhibits very weak dependence on $n$. The values of
$r_{n\ell}$ and $r_n$ converge to certain limiting values while
$n$ increases. This is due to similar $n^{-3}$ behavior of the
parameters $B_{n\ell}$ and $A_n$ at large $n$.

To estimate the limiting behavior of the ratios $R_n$ for $Z\gg1$,
note that the first term on the RHS of Eq.~(7) for $R_n$ depends on the
nuclear charge as $Z^{-2}$. Since the values of $r^f_n$ are several
times smaller than $r_n^d$ they can be neglected for $\pi\xi_Z\lesssim1$.
At these energies the second term contains only weak dependence on
$Z$.

\section{Summary}

We have considered photoionization accompanied by excitation for
helium atoms and positive two-electron ions. We focused on the
case of intermediate photon energies, for which expansion of the
amplitudes in powers of $\omega^{-1}$ is possible, while account
of the lowest term only is not sufficient. We included the final
state interactions between the electrons in the lowest order of
their Sommerfeld parameter. This enabled us to analyze the role of
various mechanisms of transferring the excitation energy.

We calculated the ratios $R_{n\ell}$ of the cross sections
$\sigma^{+*}_{n\ell}$ for ionization, accompanied by transition of
the second electron to the bound state with quantum numbers
$n,\ell$ to the cross section for ionization without excitations
$\sigma^+_{10}$, and also found the sums $R_n=\sum_\ell R_{n\ell}$
-- Eq.~(41).

Following \cite{DT}, we separated three types of contributions
beyond the high-energy limit. These are the kinematical correction
to the shake-up (SU) terms, and the contributions describing the
transfer of excitation energy by initial state and final state
interactions (ISI and FSI correspondingly). The FSI were included
by the perturbative approach developed in \cite{7} and employed in
\cite{10}. This enabled us to extract the energy dependent
factors, presenting the amplitudes in terms of the matrix elements
containing the initial state wave functions. The latter have been
obtained in \cite{8} by Correlation Function Hyperspherical
Harmonic Method. These functions were employed in atomic physics
calculations earlier \cite{9}.

We carried out the calculations, taking into account the
next-to-leading terms of expansion in powers of $\xi^2_Z$. The
kinematical corrections to the SU terms depend also on the parameter
$\pi\xi_Z$. Dependence on this parameter was included exactly.

The cross sections for excitation of $ns$ states have the most
complicated structure. In this case we had to include kinematical
corrections to SU terms. The ISI amplitudes are proportional to
$\xi^2_Z$, and we included their interference with the SU amplitudes.
The first and second order FSI amplitudes contain the factors $i\xi$
and $\xi^2$. Thus we had to include the interference between SU and FSI
amplitudes and a purely FSI term. All the corrections should be
included in the expressions for ionization cross sections without
excitation $\sigma^+_{10}$ as well. The cross section for excitation of
$p$ states was determined by ISI and FSI mechanism, with both
amplitudes being proportional to $i\xi$. Ionization accompanied by
excitations to the state with $\ell\ge2$ took place only due to the
FSI.

For the case of helium we found the values of the high energy
limits for $n\le6$ to be in agreement with those, calculated by
the others - Table~1. For excitations of $ns$ states we found the
FSI to provide the largest contributions. Excitations of $np$ and
$ns$ states, provide the contributions of the same order of
magnitude to the energy dependent parts of the ratios $R_n$.
Excitations of $nd$ states determine a corrections of about 10\%
to $R_n$. Excitation of states with $\ell\ge3$ are negligibly
small -- Tables II and III.
 For the atomic helium we carried out detailed comparison with earlier
experimental and theoretical results. We found good agreement at
$\omega\ge400\,$eV for $n=2$ and even at smaller values of $\omega$ for
the larger values of $n$ -- Fig.~2.

For larger $Z$ we found approximate formula (53), which presents
the high energy limits in the form of $Z^{-1}$ series, with the
leading terms of expansion being consistent with the earlier
results \cite{AD,AM}. Excitation of $ns$ states beyond the high
energy limit is still dominated by the FSI.  The role of
excitation of the states with $\ell=1$ increases, e.~g. for $Z=10$
transitions to $np$ states provide the largest contribution to the
energy dependent part of $R_n$.  The role of excitation of $nd$
states drops with $Z$.  These results are illustrated by
Table~VII.  In the limit $Z\gg1$ the ratios $R_n$ can be presented
as the sums of two terms. The high energy limit term does not
depend on $\omega$, dropping with $Z$ as $Z^{-2}$.  The second
term drops as $\omega^{-1}$, varying with $Z$ slowly.

For the case of helium, as well as for the ions with larger values
of $Z$, the contribution of ISI to the ratios $R_n$ is about 10\%.
Hence, the ratios $R_n$ are determined by the kinematical
corrections to SU and by the FSI.

\subsection*{Acknowledgements}

One of us (EGD) thanks for hospitality during the visit to Hebrew
University. MYaA is grateful for support by the ISF grant 134.03
and by the Hebrew University Intramural Funds,


\newpage
\subsection*{Figure captions}

Fig.1. Energy dependence of the relative role of excitations to
the states with $\ell>0$,expressed by the ratio $ X_{n
\ell}=\sigma^{+*}_{n \ell}/\sigma^{+*}_n=R_{n\ell}/R_n$ for the
case of helium ($Z=2$).

Fig.2. Energy dependence of the ratios $R_n$ in $10^{-2}$ units
for the case of helium. The dots stand for the experimental data
of \cite{1}. The solid lines show the results of the present work.
The dashed lines show the results of the calculations carried out
in \cite {IB2}.

Fig.3. Energy dependence of the relative role of excitations to
the states with $\ell>0$, expressed by the ratio $ X_{n
\ell}=\sigma^{+*}_{n \ell}/\sigma^{+*}_n=R_{n\ell}/R_n$ for the
case $Z=10$.

\newpage

\begin{table}
\caption {The high energy limits of the ratios $R_{ns}$ in percent
for helium atom. The column ``Theory-I'' \cite{Br} stands for
early high energy calculation with a variational initial state
function, ``Theory-II'' \cite{SD} results obtained using many-body
perturbation theory, Theory-III \cite{IB1} and Theory-IV
\cite{IB2} are the extensions of the low energy results with
multiconfiguration Hartree--Fock and variational ground state wave
functions correspondingly. The last column shows the result
obtained in \cite{1} by extrapolation of their experimental data
of \cite{1} with statistical errors given in parentheses}

\begin{center}
\begin{tabular}{|c|c|c|c|c|c|c|}
\hline
$n$ &Theory-I & Theory-II& Theory-III & Theory-IV&  This work& Experiment \\

\hline
$2$ &5.34 & 4.79 & 4.78& 4.79&4.80 & 4.80 (13)\\
$3$ &0.66&0.592&0.605&0.596&0.590 & 0.543 (33)\\
$4$   &0.21 &0.19 & 0.200&0.197&0.195&0.118 (37)\\
$5$ &0.100& 0.09& 0.092&0.091&0.0900&0.048 (30) \\
$6$   &0.055 & 0.05&0.050&0.0515& 0.0493&...\\
\hline
\end{tabular}
\end{center}
\end{table}

\begin{table}
\caption {Contributions of various mechanisms to the value
$B_{n0}$ defined by Eq.~(5) for the case of helium. Here
$\mu=\pi\xi_Z$; $\mu=1.04$ or $\omega=500\,$eV, $\mu=0.73$ for
$\omega=1\,$keV. }
\begin{center}
\begin{tabular}{|c|c|c|c|c|} \hline
$n$ &Kinematical corrections& ISI&FSI &$B_{n0}$\\
\hline
$2$ &$0.072(1-\mu)$&0.027&0.094&$0.193-0.072\mu$ \\
$3$ &$0.105(1-\mu)(-1)$&0.50(-2)&1.46(-2)&$(3.01-1.05\mu)(-2)$\\
$4$ &$0.036(1-\mu)(-1)$&1.76(-3)&5.00(-3)&$(1.04-0.36\mu)(-2)$\\
$5$ &$0.174(1-\mu)(-2)$&0.85(-3)&2.40(-3)&$(4.99-1.74\mu)(-3)$\\
$6$ &$0.095(1-\mu)(-2)$&0.47(-3)&1.35(-3)&$(2.77-0.95\mu)(-3)$\\
\hline
\end{tabular}
\end{center}
\end{table}

\begin{table}
\caption{Parameters $B_{n\ell}$ and $B_n$ of the energy dependent
contributions to the ratios $R_{n\ell}$ and $R_n$ defined by Eqs.
(5)--(7).}

\begin{center}
\begin{tabular}{|c|c|c|}
\hline
States & $B_{n\ell}$ & $B_n$\\
\hline
$2s$ & $0.193-0.072\mu$ & \\
$2p$ &  0.130 & $0.323-0.072\mu$\\
\hline
$3s$ & $(3.01-1.05\mu)(-2)$ &\\
$3p$ &  $1.86(-2)$ & $(5.13-1.05\mu)(-2)$\\
$3d$ &  $3.07(-3)$ & \\
\hline
$4s$ & $(1.04-0.36\mu)(-2)$ & \\
$4p$ & $0.62(-2)$ & $(1.80-0.36\mu)(-2)$ \\
$4d$ & $1.35(-3)$ & \\
$4f$ & $3.9(-5)$ & \\
\hline
$5s$ & $(4.99-1.74\mu)(-3)$ & \\
$5p$ & $2.93(-3)$ & $(0.86-0.17\mu)(-2)$ \\
$5d$ & $0.70(-3)$ & \\
\hline
$6s$ & $(2.77-0.95\mu)(-3)$ & \\
$6p$ & $1.63(-3)$ & $(4.80-0.95\mu)(-3)$ \\
$6d$ & $0.40(-4)$ & \\
\hline
\end{tabular}
\end{center}
\end{table}

\begin{table}
\caption
{The values $A_nZ^2\cdot10^2$, with $A_n$ defined by Eq.~(9).}
\begin{center}
\begin{tabular}{|c|c|c|c|c|c|}
\hline
$n$ & $Z=2$ & $Z=3$ & $Z=4$ & $Z=6$ & $Z=10$ \\
\hline
$2$ &19.1& 14.9& 12.8& 11.4&10.4 \\
$3$ &2.36&2.18&2.06&1.93&1.84 \\
$4$ &0.781&0.749& 0.722&0.692&0.660\\
$5$ &0.360& 0.351& 0.340&0.327& 0.316\\
$6$ &0.197&0.193&0.188&0.182&0.176\\
\hline
\end{tabular}
\end{center}
\end{table}

\begin{table}
\caption
{The values of coefficients $a_n$ and $b_n$ in Eq.~(53) and of
coefficients $c_n$ in Eq.~(54).}

\begin{center}
\begin{tabular}{|c|c|c|c|}
\hline
$n$ &$a_n(-2)$ & $b_n(-2)$ & $c_n(-2)$ \\
\hline
$2$ &8.9 & 15.0 & 9.2 \\
$3$ &1.7&1.4 & 1.7\\
$4$ &0.61&0.48 &0.64\\
$5$ &0.30& 0.16 & 0.30\\
$6$ &0.17&0.08 & 0.17\\
\hline
\end{tabular}
\end{center}
\end{table}

\begin{table}
\caption
{Contributions of various mechanisms to
the value $B_n$ defined by Eq.~(5) for the case $Z=10$, $\mu=\pi\xi_Z$.}
\begin{center}
\begin{tabular}{|c|c|c|c|c|}
\hline
$n$ &Kinematical corrections& ISI & FSI & $B_{n0}$ \\
\hline
2 & $ 0.390(1-\mu)(-1)$ & $1.91(-2)$ & $4.96(-2)$ & $0.104-0.039\mu$\\

3 & $0.820(1-\mu)(-2)$ & $0.41(-3)$ & $1.00(-2)$ &
$(2.23-0.82\mu)(-2)$\\

4 & $0.326(1-\mu)(-2)$ & $1.56(-3)$ & $3.85(-3)$ &
$(8.67-3.26\mu)(-3)$\\

5 & $0.152(1-\mu)(-2)$ & $0.76(-3)$ & $1.91(-3)$ &
$(4.19-1.52\mu)(-3)$ \\

6 & $0.856(1-\mu)(-2)$ & $0.43(-3)$ & $1.07(-3)$ &
$(2.36-0.86\mu)(-3)$\\
\hline
\end{tabular}
\end{center}
\end{table}

\begin{table}
\caption {The values of characteristics $r^d_{n0}$ for $\ell=0$ and
$r_{n\ell}$ for $\ell>0$  and $r^d_n$ defined by Eqs.
(58), (55) and (56) for $Z=2$ and $Z=10$.}

\begin{center}
\begin{tabular}{|c|c|c|c|c|}
\hline
 & \multicolumn{2}{c|}{$Z=2$} & \multicolumn{2}{c|}{$Z=10$} \\
\hline

State & $r_{n0}^d,r_{n\ell}$ & $ r^d_n$ & $r^d_{n0},r_{n\ell}$ & $r^d_n$\\
\hline

$2s$ & 1.01 & & 1.00 & \\

$2p$ & 0.68  & 1.69  & 1.38  & 2.38\\
\hline

$3s$   & 1.28 & & 1.21 & \\

$3p$ & 0.77 & 2.18  & 1.68  & 2.94 \\

$3d$   & 0.13  &   & 0.05 & \\
\hline

$4s$ & 1.33 & & 1.31 & \\
$4p$ & 0.80 & 2.33 & 1.80  & 3.18 \\
$4d$ & 0.17 & & 0.07 & \\
$4f$ & 5.0(-3) & & $2.0(-3)$ & \\
\hline

$5s$ & 1.39 & & 1.33 & \\
$5p$ & 0.82 &2.40 & 1.86 & 3.27 \\
$5d$ & 0.19 & & 0.08 & \\
\hline

$6s$ & 1.40& &      1.34 & \\
$6p$ & 0.83 & 2.43 & 1.89 & 3.31\\
$6d$ & 0.20 & & 0.08 & \\
\hline

\end{tabular}
\end{center}
\end{table}

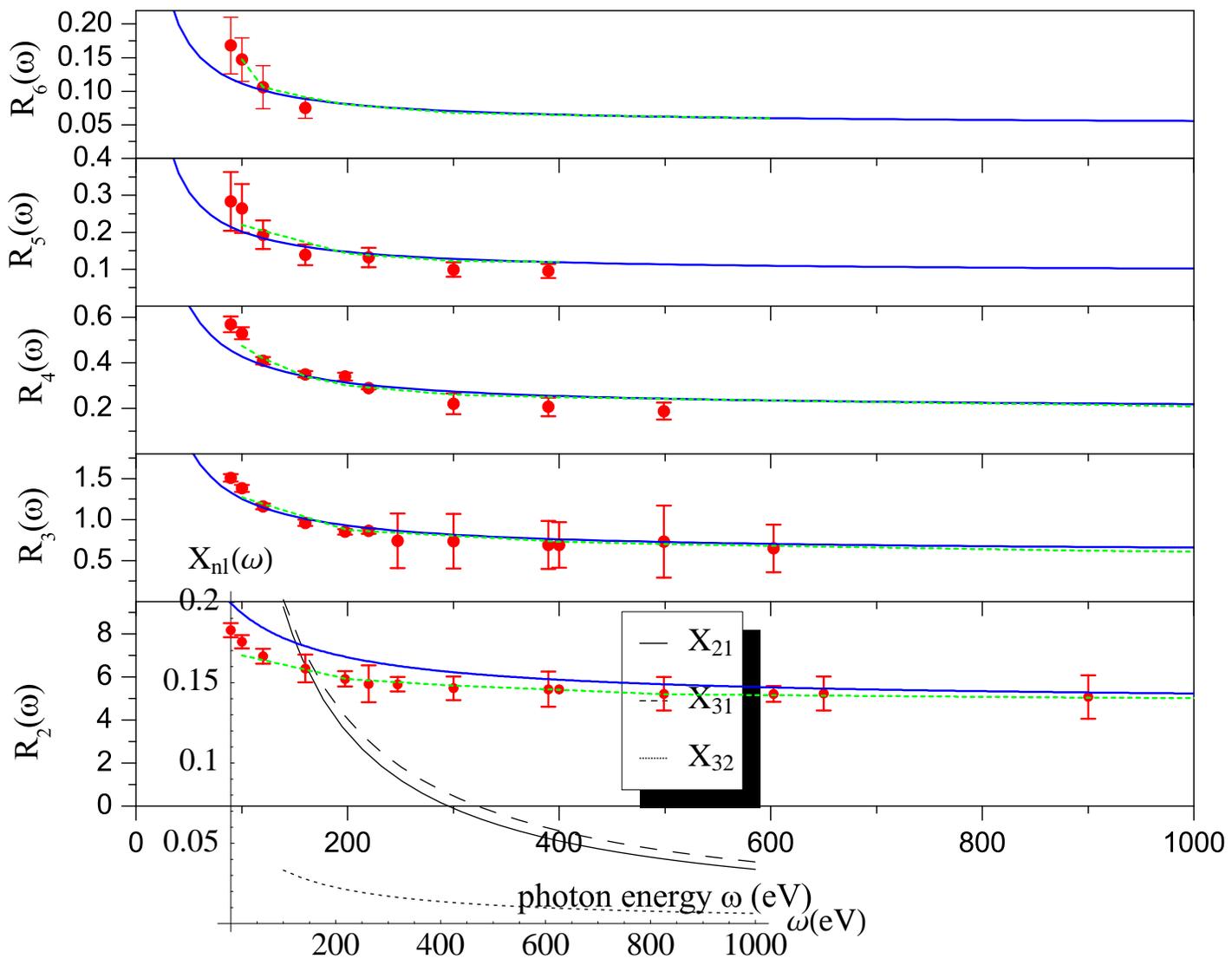
\end{document}